\documentclass[11pt]{article}
\usepackage{jheppub}
\usepackage{epstopdf}
\usepackage{enumitem}
\usepackage{tcolorbox}

\usepackage{jheppub}
\usepackage{mathrsfs}
\usepackage{psfrag}
\usepackage{color}
\usepackage{hyperref}

\usepackage{slashed}
\usepackage{feynmp-auto}
\usepackage{simplewick}
\usepackage{cancel}

\usepackage{todonotes}

\usepackage{amsmath}
\usepackage{amsfonts}
\usepackage{graphicx}
\usepackage{amssymb}
\usepackage{xcolor}

\usepackage{mathtools}

\usepackage{amsmath,bbm,array,amsfonts,graphicx,wrapfig,arydshln,lscape,float,multirow,longtable,rotating,makecell}
\usepackage{url}

\DeclareMathAlphabet\mathbfcal{OMS}{cmsy}{b}{n}

\newcommand{\ra}{\rangle}

\newcommand{\p}{\partial}

\newcommand{\la}{\langle}

\newcommand{\eq}{\begin{equation}}
\newcommand{\eqe}{\end{equation}}
\newcommand{\eqa}{\begin{eqnarray}}
\newcommand{\eqae}{\end{eqnarray}}

\newcommand{\bn}{\begin{enumerate}}
\newcommand{\en}{\end{enumerate}}

\newcommand{\eqc}[1]{(\ref{#1})}

\def\CO{{\mathcal O}}

\def\CO{{\mathcal O}}

\def\a{\alpha}
\def\b{\beta}
\def\g{\gamma}
\def\e{\epsilon}

\def\z{\zeta}

\def\k{\kappa}
\def\l{\lambda}
\def\m{\mu}
\def\n{\nu}

\def\s{\sigma}

\def\G{\Gamma}

\def\S{\Sigma}

\def\half{\frac{1}{2}}

\def\p{\partial}

\newcommand{\bfig}{\begin{figure}}
\newcommand{\efig}{\end{figure}}

\def\bl#1\el{\begin{align} #1 \end{align}}
\def\bg#1\eg{\begin{gather} #1 \end{gather}}

\def\bld#1\eld{\begin{aligned} #1 \end{aligned}}
\def\bgd#1\egd{\begin{gathered} #1 \end{gathered}}

\newcommand{\bra}[1]{\langle{#1}|}
\newcommand{\ket}[1]{|{#1}\rangle}

\newcommand{\sbra}[1]{ [{#1} |}
\newcommand{\sket}[1]{ | {#1} ]}

\renewcommand{\tt}{\texttt}
\renewcommand{\bf}{\textbf}

\newcommand{\AB}[1]{\langle #1 \rangle}
\newcommand{\SB}[1]{[ #1 ]}
\newcommand{\MixLeft}[3]{\langle #1 | #2 | #3 ]}
\newcommand{\BS}[1]{\boldsymbol{#1}}
\newcommand{\RAB}[1]{| #1 \rangle}
\newcommand{\LAB}[1]{\langle #1 |}
\newcommand{\RSB}[1]{| #1 ]}
\newcommand{\LSB}[1]{[ #1 |}

\newcommand{\RN}[1]{%
  \textup{\uppercase\expandafter{\romannumeral#1}}%
}

\newcommand*{\mathcolor}{}
\def\mathcolor#1#{\mathcoloraux{#1}}
\newcommand*{\mathcoloraux}[3]{%
  \protect\leavevmode
  \begingroup
    \color#1{#2}#3%
  \endgroup
}

\title{Kerr-Newman stress-tensor from minimal coupling}
\author{Ming-Zhi Chung$^{1}$}
\author{Yu-tin Huang$^{1,2}$}
\author{Jung-Wook Kim$^{3}$}

\affiliation{$^1$ Department of Physics and Astronomy, National Taiwan University, Taipei 10617, Taiwan}
\affiliation{$^2$ Physics Division, National Center for Theoretical Sciences, National Tsing-Hua University, No.101, Section 2, Kuang-Fu Road, Hsinchu, Taiwan}
\affiliation{$^3$ Department of Physics and Astronomy, Seoul National University, Seoul 08826, Korea
}

\emailAdd{dchung0741@gmail.com}
\emailAdd{yutinyt@gmail.com}
\emailAdd{jwkonline@snu.ac.kr}

\abstract{In this paper, we demonstrate that at leading order in post Minkowskian (PM) expansion, the stress-energy tensor of Kerr-Newman black hole can be recovered to all orders in spin from three sets of minimal coupling: the electric and gravitational minimal coupling for higher-spin particles, and the ``minimal coupling" for massive spin-2 decay. These couplings are uniquely defined from kinematic consideration alone. This is shown by extracting the classical piece of the one-loop stress-energy tensor form factor, which we provide a basis that is valid to all orders in spin. The 1 PM stress tensor, and the metric in the harmonic gauge, is then recovered from the classical spin limit of the form factor.}

\begin{document}
\begin{flushright}
\vspace{10pt} \hfill{NCTS-TH/1910} \vspace{20mm}
\end{flushright}
\maketitle

\section{Introduction and conclusion}\label{sec:Intro}
For the past few years,  a new facet of the simplicity of four-dimensional black hole like spacetimes has been uncovered through the study of its image in scattering amplitudes. By black hole like we refer to classical solutions that are completely characterized by a finite set of charges. This new ``on-shell" point of view elects to begin with the physical observables associated with non-trivial spacetimes such as the classical impulse and the scattering angle. In the post-Minkowskian expansion, or perturbation in Newton's constant $G$, these observables can be encoded in the null transfer momenta limit ($q^2\rightarrow 0$) of the four-point scattering amplitude~\cite{Bjerrum-Bohr:2014zsa, Kosower:2018adc, Maybee:2019jus}, where $q$ is the transverse momenta. Importantly, the ``particle" for the scattering amplitude share the same quantum numbers as the classical solution. 

While this point of view has been long appreciated for Schwarzschild black holes/scalar particles, only recently has it been shown that to all order in spins, the effective stress-energy tensor for Kerr black holes is given by the classical-spin limit, $s\rightarrow \infty$ $\hbar\rightarrow 0$ while $s\hbar$ held fixed, of minimally coupled spinning particles~\cite{Guevara:2017csg, Chung:2018kqs,Guevara:2018wpp,Arkani-Hamed:2019ymq}. While there are no known elementary higher-spin particles in nature, these minimal couplings can be defined in a purely kinematic fashion~\cite{Arkani-Hamed:2017jhn}. Further introducing a complex phase to the coupling the corresponds to a gravitational  duality rotation, resulting in that of Taub-NUT spacetime~\cite{Huang:2019cja}. The fact that the simplest three-point amplitude encodes the leading post Minkowskian (1 PM), or leading $G$ order, stress-energy tensor of black hole solutions, can be viewed as an on-shell statement of the no hair theorem.

\begin{figure}
\begin{center}
\includegraphics[scale=0.45]{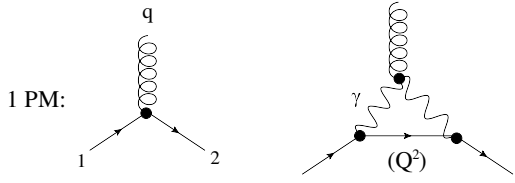}
\caption{Contributions to the 1 PM stress tensor form factor. The purely gravitational contribution is given by the tree-level gravitational minimal coupling, while the photon couplings contribute through one-loop effects. Since the charge appears as $Q^2G$, the one-loop diagram is still 1 PM.  }
\label{Fig1}
\end{center}
\end{figure}

In the presence of extra massless fields, the stress-energy tensor will be modified at 1 PM. These modifications come in as one-loop corrections to the stress tensor form factor, as shown in fig. \ref{Fig1}, reflecting the fact that the object is charged under the additional massless fields. Note that while this is a loop effect, it retains a classical piece. This is because after restoring $\hbar$, the mass appears in the propagator in the form $m \hbar^{-1}$, so a factor of $\hbar$ from an additional loop can be cancelled by an additional factor of $m \hbar^{-1}$~\cite{Holstein:2004dn}
. Indeed such a phenomenon was pointed out long ago in the computation of the one-loop stress tensor form factor in scalar QED~\cite{Donoghue:2001qc,Holstein:2004dn}. There, the form factor is expanded as:
\bl
\bra{p_2} T_{\m\n} (q) \ket{p_1} &= \frac{1}{\sqrt{4 E_1 E_2}} \left[ 2 P_\m P_\n F_1 (q^2) + (q_\m q_\n - \eta_{\m\n} q^2 ) F_2(q^2) \right]\, \label{eq:TFF}
\el
where $P=\frac{p_1{+}p_2}{2}$ and $q=p_2{-}p_1$, and conservation is manfest. It was shown that the one-loop diagram in fig.\ref{Fig1} indeed leads to non-trivial corrections to the scalar functions $F_1 (q^2)$ and $F_2 (q^2)$ in the classical limit $\hbar \to 0$. This term has a classical field theory interpretation as well, which becomes manifest in the Breit frame; the charged scalar particle at the origin sources electromagnetic fields, and this in turn contributes to the stress tensor. The same analysis can be extended to spinning particles as well. It has been shown in~\cite{Donoghue:2001qc,Holstein:2006ud} that the same computation for spin-$\half$ and spin-1 particles generate contributions to the form factor eq.\eqc{eq:TFF} which is linear in spin variables, and this spin-linear contribution matches to the corresponding spin order contributions to the stress tensor of electromagnetic fields generated by spinning charged black holes(BH), also known as Kerr-Newman BHs.

More recently~\cite{Moynihan:2019bor}, the relation between Kerr-Newman and minimal coupling has been tested up to spin-1 via the computation of the classical potential utilizing the on-shell formulation introduced in~\cite{Arkani-Hamed:2017jhn}. However, similar to~\cite{Chung:2018kqs}, due to the lack of prescription for Compton scattering beyond spin-1, the classical potential was only computed up to degree 2.

In this paper, we approach the stress tensor form factor from a completely on-shell viewpoint. To begin, we first demonstrate that the current form factor induced by the electromagnetic gauge potential of the Kerr-Newman black hole is indeed given by higher-spin particles minimally coupled to photons. This is done by deriving the one-particle world-line effective action of a spinning particle from the electromagnetic background of the Kerr-Newman black hole solution, and demonstrating that the three-point amplitude constructed from this world-line effective action matches with that of minimal coupling in the classical-spin limit. Similar analysis was done for the comparison of gravitational minimal coupling and that of Kerr black hole~\cite{Chung:2018kqs}. 

After identifying the electromagnetic coupling, we proceed to compute the classical piece of the one-loop form factor at 1 PM via unitarity methods~\cite{Forde:2007mi}. Since we are still at 1 PM, the loop under discussion is that of photons. Note that in the original computation for spin-$\frac{1}{2}$ and spin-$1$ ~\cite{Donoghue:2001qc,Holstein:2006ud}, the basis for the stress tensor was constructed out of momenta and external wave-functions of the spinning-particle. In other words, the basis is constructed in a case by case manner with distinction between half-integer and integer spins. Here, instead, we introduce a universal basis for which taking the classical-spin limit is straightforward,
\begin{equation}\label{SpinBasis}
\bra{p_2} T_{\m\n} (q) \ket{p_1} =\frac{1}{\sqrt{4 E_1 E_2}} \left[  F_1 P_{\mu}P_{\nu} + 2 F_2 P_{(\mu}E_{\nu)} + F_3(q_{\mu}q_{\nu} - \eta_{\mu\nu} q^2) + F_4 E_{\mu}E_{\nu}\right]
\end{equation}
where $E_\mu$ is defined in eq.(\ref{EDef}). The relevant classical piece lies in the scalar triangle integral with two massless internal propagators one one massive. It's coefficient can be computed from the triple-cut, which is given by the product of two electric minimal coupling and the stress-tensor form factor with photons as external states, i.e the Maxwell stress-tensor:
\eq\label{eq:On-shell_amplitude}
\vcenter{\hbox{\includegraphics[scale=0.45]{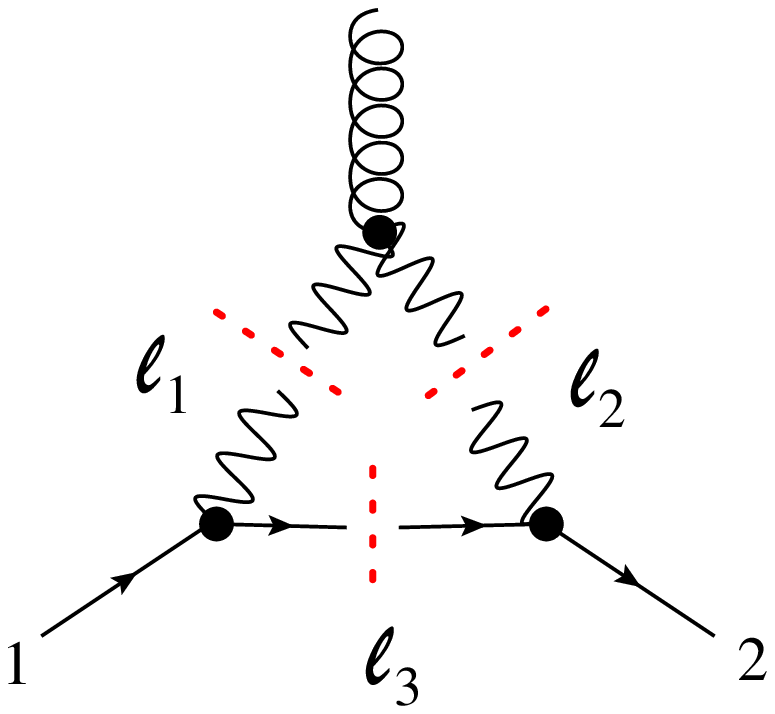}}}\;= \bra{\gamma_{\ell_1}} T_{\m\n} (q) \ket{\gamma_{\ell_2}}\otimes A_3(1^s\ell_1\ell^s_3)\otimes A_3(\ell^s_3\ell_22^s)\,.
\eqe  
Note that each of these three-point couplings can be uniquely determined kinematically. \textit{Importantly, we take the classical-spin limit directly on the cut}. As a result, the two minimal couplings exponentiate, as discussed in~\cite{Arkani-Hamed:2017jhn}, and the extraction of triangle coefficient is greatly simplified. Utilising the universal basis for the stress-tensor form factor we have presented, we derive the coefficients for these bases at leading order in $q^2\rightarrow 0$ limit. The result is then compared to the photon stress-energy tensor, $T_{\mu\nu}= - F_{\mu\sigma}F^{\sigma}\,_{\nu} + \frac{1}{4}\eta_{\nu\nu}F^{\rho\sigma}F_{\rho\sigma}$, evaluated on the electromagnetic potential of Kerr-Newman black hole, and we find complete agreement. This verifies that the set of minimal coupling completely reproduces the stress-tensor form factor to all orders in spin.

This paper is organized as follows: in the next section we first derive the electro-magnetic part of the one-particle EFT for Kerr-Newman black holes, and match the resulting three-point amplitude to that of minimal coupling. Then for later purpose we also derive the corresponding Maxwell stress-tensor, and cast it in our preferred kinematic basis. In sec.3, we reproduce the 1 PM form factor at $\mathcal{O}(Q^2)$ via unitarity methods, with special focus on taking the classcial-spin limit on the cut.

\newpage
\section{EM minimal coupling and the Kerr-Newman one-particle EFT}
 The one-particle effective action aims to give an effective description of the gravitational or electromagnetic coupling for a compact source. Setup as a world-line action where the world-line degrees of freedom couple to external background fields through a (infinite) series of local operators~\cite{Goldberger:2004jt}, their associated Wilson coefficients then encode the effective stress-tensor/current that sources the linearized gravitational/electromagentic field. For spinning objects, the world-line theory attains extra spin fields~\cite{Porto:2005ac}, which introduce an infinite number of multipole moments each with its own Wilson coefficient~\cite{Porto:2008jj}, (see also~\cite{Levi:2015msa}).

In this section, we determine the 1 PM effective action for Kerr-Newman black hole. Since the black hole is charged, it will source both gravitational and electromagnetic fields. To leading order in $G$, the $Q$ independent part of the effective stress tensor is identical to that of Kerr, whose form can be found in~\cite{Vines:2017hyw}. Here we will be interested in the effective current, through which the world-line couples to the Maxwell field. 

We begin by reviewing the Kerr-Schild form of the Kerr-Newman solution. Via suitable gauge transformation, just as the Kerr-metric, the vector potential for Kerr-Newman can be written as an differential operator acting on the vector potential of a point source in flat space. As the latter simply corresponds to a point charge on the world-line minimally coupled to the Maxwell field, $e u^\mu A_\mu$, one can straightforwardly obtain the Kerr-Newman world-line action. We compute the three-point amplitude from this action and demonstrate that it matches with minimal coupling of  spinning particles. This is analogous to the situation for Kerr BHs~\cite{Chung:2018kqs,Chung:2019duq}. Note that in a sense this result is expected from double copy:  in the Kerr-Schild form, the single copy of the Kerr solution is the Kerr-Newman vector potential~\cite{Monteiro:2014cda}, while for the on-shell side, the single copy of gravitational minimal coupling is the photon minimal coupling.

For later purpose, we derive the stress tensor induced by the electromagnetic field sourced by the Kerr-Newman BH, which is the matter stress tensor that enters in the Einstein's equation $G_{\m\n} = - 8 \pi G T_{\m\n}$. The expression will be later matched to a corresponding computation using QFT techniques. 

\subsection{The one-particle EFT for Kerr-Newman}\label{subsec:Photon_Effective_Action}
Kerr-Schild form is a representation for solutions to Einstein's equations which reduces the non-linearities of GR to linear differential equations.\footnote{For a review, consult e.g.~\cite{Stephani:2003tm}.} It is precisely in this form that the double copy relations between classical solutions for Einstein gravity and linearized Yang-Mills was found~\cite{Monteiro:2014cda}. 

The Kerr-Newman solution in Kerr-Schild form is given by\footnote{Following usual conventions of GR literature, we use Gaussian units in this section.}
\bl
\bld
g_{\m\n} &= \eta_{\m\n} - f k_\m k_\n
\\ f &= \frac{G r^2}{r^4 + a^2 z^2} [2 Mr - Q^2]
\\ k_\m &= \left( 1, \frac{rx + ay}{r^2 + a^2}, \frac{ry - ax}{r^2 + a^2}, \frac{z}{r} \right)
\\ A_\m &= \frac{Q r^3}{r^4 + a^2 z^2} k_\m
\eld \label{eq:KNBHinKSform}
\el
where $k^\m$ is null both in $\eta_{\m\n}$ and $g_{\m\n}$, and $r$ is implicitly defined by the relation
\bl
1 &= \frac{x^2 + y^2}{r^2 + a^2} + \frac{z^2}{r^2} \,.
\el
To simplify the expressions, we adopt the following definitions.
\bl
\bgd
x = R^1 \,,\quad y = R^2 \,,\quad z = R^3 \,,
\\ R = \sqrt{x^2 + y^2 + z^2} \,,
\\ \S = \sqrt{(x^2+y^2+z^2-a^2)^2 + 4 a^2 z^2} = \frac{r^4 + a^2 z^2}{r^2} \,.
\egd
\el
When computing the moments of the source, the following relations given in~\cite{Vines:2017hyw} will prove to be useful.
\bl
\bld
\frac{r}{\S} &= \cos (\vec{a} \cdot \vec{\nabla}) \frac{1}{R}
\\ \frac{r}{\S} \frac{\vec{R} \times \vec{a}}{r^2 + a^2} &= \sinh(\vec{a} \times \vec{\nabla}) \frac{1}{R} \,.
\eld \label{eq:sourcematching}
\el


In eq.\eqc{eq:KNBHinKSform} the vector potential $A_\m$ is given as
\bl
A_\m &= \frac{Q r}{\S} k_\m
\el
which does not satisfy the Lorenz gauge condition $\nabla^\m A_\m = 0$. The gauge connection $A'_\m dx^\m$ that satisfies the Lorenz gauge condition can be obtained from $A_\m dx^\m$ by the following gauge transformation
\bl
\bld
A'_\m dx^\m &= A_\m dx^\m - \frac{Q}{2} d \log (r^2 + a^2) 
\\ &= \frac{Qr}{\S} \left[ dt + \frac{a}{a^2 + r^2} \left( y dx - x dy \right) \right]
\eld
\el
which, in dual representation, is
\bl
A'^\m \p_\m &= \frac{Qr}{\S} \left[ \p_t - \frac{a}{a^2 + r^2} \left( y \p_x - x \p_y \right) \right] = \left( \frac{Qr}{\S}, - \frac{Qr}{\S} \frac{\vec{R} \times \vec{a}}{r^2 + a^2} \right) \,. \label{eq:LinEM}
\el
Using eq.\eqc{eq:sourcematching}, the linearised gauge connection in Lorenz gauge eq.\eqc{eq:LinEM} can be written as
\bl
A'^\m \p_\m &= \left( \cos(\vec{a} \cdot \vec{\nabla}) , - \sinh (\vec{a} \times \vec{\nabla}) \right) \frac{Q}{R} \label{eq:MaxwellSol}
\el
which can be covariantised by $u^\m \p_\m = \p_t$~\cite{Vines:2017hyw}
\bl
A'^\m &= \left( u^\m \cos(a \cdot \p) + \e^{\m\n\a\b} u_\n a_\a \p_\b \frac{\sin(a \cdot \p)}{a \cdot \p} \right) \frac{Q}{R} \,.
\el

From the above, one can determine the effective current on the world-line $j^\m$ as
\bl
\bld
j^\m &= Q \int ds \left[ u^\m \cos(a \cdot \p) + \e^{\m\n\a\b} u_\n a_\a \p_\b \frac{\sin(a \cdot \p)}{a \cdot \p} \right] \delta^4 \left[ x - x_{\text{wl}}(s) \right]
\\ &= Q \int ds \sum_{n=0}^{\infty} \left[ u^\m \frac{\left( - (a \cdot \p)^2 \right)^n}{(2n)!} + \e^{\m\n\a\b} u_\n a_\a \p_\b \frac{\left( - (a \cdot \p)^2 \right)^n}{(2n+1)!} \right] \delta^4 \left[ x - x_{\text{wl}}(s) \right]\,.
\eld
\el
This implies that the world-line couples through the background gauge field as
\bl
S_{int} &= - 4 \pi \int d^4 x A_\m j^\m \,.
\el
 The derivatives on the source can be replaced by derivatives on the gauge field $A_\m$ through integration by parts.
\bl
\bld
S_{int} &= - \int d^4 x ~4 \pi Q \int ds ~\delta^4 \left[ x - x_{\text{wl}}(s) \right]
\\ &\phantom{=asdfasdfasdf} \times \sum_{n=0}^{\infty} \left[ u^\m \frac{\left( - (a \cdot \p)^2 \right)^n}{(2n)!} - \e^{\m\n\a\b} u_\n a_\a \p_\b \frac{\left( - (a \cdot \p)^2 \right)^n}{(2n+1)!} \right] A_\m (x)
\eld \label{eq:3ptIntVert}
\el
Indeed taking $a\rightarrow 0$ one recovers the minimal electromagnetic coupling in flat space. 

From the above action, we can extract the EFT three-point amplitude by substituting the polarisation tensors, e.g. $A^\m \to \e^\m$~\cite{Chung:2018kqs,Chung:2019duq}. Taking all momenta to be incoming, labeling the photon as $p_3$ and $u^\m = \frac{p_1^\m - p_2^\m}{2m}$, the operators can be mapped to:
\bl
\bld\label{eq:Element}
u^\m A_\m &\to \frac{x^{\eta}}{\sqrt{2}}
\\ - \e^{\m\n\a\b} u_\n a_\a \p_\b A_\m &\to \frac{x^{\eta}}{\sqrt{2}} \left( - \eta \frac{p_3 \cdot S}{m} \right)
\\ - (a \cdot \p)^2 &\to \left( \frac{p_3 \cdot S}{m} \right)^2 = \left( - \eta \frac{p_3 \cdot S}{m} \right)^2
\eld
\el
where $x$ is defined by the kinematics $2 q\cdot p_1 = \MixLeft{q}{p_1}{q} = 0$ \citep{Arkani-Hamed:2017jhn}
\begin{equation}
x \lambda_{q}^{\alpha} = \tilde{\lambda}_{q, \dot{\alpha}}\frac{p_1^{\dot{\alpha}\alpha}}{m}
\end{equation}
and the spin operator $S$ is defined by the Pauli-Lubanski vector. The chiral and anti-chiral representation of a spin operator contracted with $q$ are: \citep{Chung:2018kqs, Chung:2019duq}
\begin{equation}\label{eq:Chiral Anti-Chiral Spin}
\begin{split}
(q \cdot S)_{\alpha_1 \cdots \alpha_{2s}}^{\phantom{a}\beta_1 \cdots \beta_{2s}} 
&= \sum_{i=1}^{2s} + \frac{x}{2}\left(\RAB{q}\LAB{q}\right)_{\alpha_i}^{\phantom{a} \beta_i} \mathbb{I}_i \stackrel{\cdot}{=} +(2s)\frac{x}{2}\left(\RAB{q}\LAB{q}\right)_{\alpha_1}^{\phantom{a} \beta_1} \mathbb{I}_1\\
(q \cdot S)^{\dot{\alpha}_1 \cdots \dot{\alpha}_{2s}}_{\phantom{a}\dot{\beta}_1 \cdots \dot{\beta}_{2s}} 
&= \sum_{i=1}^{2s} - \frac{1}{2x} \left(\RSB{q}\LSB{q}\right)^{\dot{\alpha}_i}_{\phantom{a} \dot{\beta}_i} \bar{\mathbb{I}}_i \stackrel{\cdot}{=} -(2s)\frac{1}{2x} \left(\RSB{q}\LSB{q}\right)^{\dot{\alpha}_1}_{\phantom{a} \dot{\beta}_1} \bar{\mathbb{I}}_1 \,,
\end{split}
\end{equation}
where $\stackrel{\cdot}{=}$ incorporates symmetrisation of the massive spinors. In the symmetric basis,\footnote{
In eq.\eqref{eq:Chiral Anti-Chiral Spin}, the identity operators are defined as
$$
\mathbb{I}_i = \delta_{\alpha_1}^{\beta_1} \cdots \delta_{\alpha_{i-1}}^{\beta_{i-1}} \delta_{\alpha_{i+1}}^{\beta_{i+1}} \cdots \delta_{\alpha_{2s}}^{\beta_{2s}}, \quad 
\bar{\mathbb{I}}_i = \bar{\delta}^{\dot{\alpha}_1}_{\dot{\beta}_1} \cdots \bar{\delta}^{\dot{\alpha}_{i-1}}_{\dot{\beta}_{i-1}} \bar{\delta}^{\dot{\alpha}_{i+1}}_{\dot{\beta}_{i+1}} \cdots \bar{\delta}^{\dot{\alpha}_{2s}}_{\dot{\beta}_{2s}}
$$, in eq.\eqref{eq:Symmetric Spin}, the identity operators are definded as
$$
\mathbb{I} = \delta_{\alpha_1}^{\beta_1} \cdots \delta_{\alpha_{s}}^{\beta_{s}}, \quad 
\bar{\mathbb{I}} = \bar{\delta}^{\dot{\alpha}_1}_{\dot{\beta}_1} \cdots \bar{\delta}^{\dot{\alpha}_{s}}_{\dot{\beta}_{s}}
$$
}
\begin{equation}\label{eq:Symmetric Spin}
(q \cdot S)^{\dot{\alpha}_1 \cdots \dot{\alpha_s} \phantom{abc} \beta_1 \cdots \beta_s}_{\phantom{a} \dot{\beta}_1 \cdots \dot{\beta_s}, \alpha_1 \cdots \alpha_s} 
\equiv 
(q \cdot S)^{\lbrace\dot{\alpha}_s\rbrace\phantom{ab} \lbrace\beta_s\rbrace}_{\phantom{a}\lbrace \dot{\beta}_s\rbrace \lbrace\alpha_s\rbrace}
= 
(q \cdot S)_{\alpha_1 \cdots \alpha_{s}}^{\phantom{a}\beta_1 \cdots \beta_{s}}\bar{\mathbb{I}} + (q \cdot S)^{\dot{\alpha}_1 \cdots \dot{\alpha}_{s}}_{\phantom{a}\dot{\beta}_1 \cdots \dot{\beta}_{s}} \mathbb{I}
\end{equation}
Summing up all the elements in \eqref{eq:Element}, the three-point amplitude becomes
\bl
M_s^{\eta} = \e^\ast(\bf{2})^{\{\dot{\beta}_s\} \{\beta_s\}} \left[ \frac{4 \pi Q x^{\eta} }{\sqrt{2}} \sum_{n=0}^{\infty} \frac{1}{n!} \left( - \eta \frac{q \cdot S}{m} \right)^n \right]_{\{\beta_s\} \{\dot{\beta}_s\}}^{\{\dot{\alpha}_s\}\{\alpha_s\}} \e(\bf{1})_{\{\alpha_s\}\{\dot{\alpha}_s\}} \label{eq:EM3ptEFT}
\el
which has a very similar structure to EFT graviton three-point coupling~\cite{Chung:2018kqs,Chung:2019duq}
\bl
M_s^{2 \eta} = \e^\ast(\bf{2}) \left[ \frac{\k m x^{2\eta} }{2} \sum_{n=0}^{\infty} \frac{C_{\text{S}^n}}{n!} \left( - \eta \frac{q \cdot S}{m} \right)^n \right] \e(\bf{1}) \,. \label{eq:Grav3ptEFT}\,.
\el

The three-point amplitude for electro-magnetic, and gravitational minimal coupling of a spin-$s$ particle is given as:
\eq\label{MinimalAmp}
\mathcal{A}_3^{\pm}=\sqrt{2}e mx^{\pm}\left(\frac{\langle \mathbf{12}\rangle}{m}\right)^{2s},\quad\mathcal{M}_3^{\pm}=\frac{\kappa}{2}m^2x^{\pm2}\left(\frac{\langle \mathbf{12}\rangle}{m}\right)^{2s}\,,
\eqe
respectively. Now, it is known that Kerr BHs have Wilson coefficients $C_{\text{S}^n} = 1$ in eq.\eqc{eq:Grav3ptEFT}, and matches to gravitational minimal coupling in eq.(\ref{MinimalAmp}) in the classical-spin limit $s \to \infty$~\cite{Chung:2018kqs}. The same conclusion can then be drawn for the amplitude eq.\eqc{eq:EM3ptEFT} and electro-magnetic amplitude in eq.(\ref{MinimalAmp}), thus \textit{electro-magnetic minimal coupling for spinning particles, defined in eq.(\ref{MinimalAmp}), will reduce to Kerr-Newman BHs in the $s \to \infty$ limit\footnote{For finite $s$ results, consult ref.\cite{Chung:2019duq}.}.}

\subsection{Stress tensor of Kerr-Newman solution}
The Kerr-Newman BH sources electromagnetic fields that contribute to the stress tensor. This contribution is given by the electromagnetic stress tensor, which is defined as
\bl
T_{\m\n} &= - \frac{1}{4 \pi} F_{\m\l} F_{\n}^{~\l} + \frac{1}{16 \pi} \eta_{\m\n} F_{\a\b} F^{\a\b} \,. \label{eq:EMstress}
\el
To compute this quantity, we need to determine the electromagnetic fields that the BH has generated. The computations are easier with vector calculus for 3d Euclidean space, so we will fix the frame to be the rest frame of the BH where BH is at the origin. From eq.\eqc{eq:MaxwellSol} we may deduce the electric and magnetic fields as follows.
\bl
\bld
\vec{E} &= - \vec\nabla \cos(\vec{a} \cdot \vec{\nabla}) \frac{Q}{R}
\\ \vec{B} &= - \vec\nabla \times \sinh (\vec{a} \times \vec{\nabla}) \frac{Q}{R}
\eld
\el
Since $\frac{1}{R}$ is harmonic, $\vec\nabla \times ( \vec{a} \times \vec{\nabla})$ can be considered as $- \vec\nabla (\vec{a} \cdot \vec{\nabla})$ in the above expression. Therefore an equivalent representation for the magnetic field is
\bl
\vec{B} &= \vec\nabla \sin(\vec{a} \cdot \vec{\nabla}) \frac{Q}{R}
\el
This motivates the following definition of holomorphic(anti-holomorphic) electromagnetic fields $\vec{H}$($\vec{\bar{H}}$);
\bl
\bld
\vec{H} &= \vec{E} - i \vec{B} = - \vec\nabla e^{i \vec{a} \cdot \vec{\nabla}} \frac{Q}{R} = - \vec\nabla \frac{Q}{\sqrt{x^2 + y^2 + (z+ia)^2}} = - \vec\nabla Q f(R)
\\ \vec{\bar{H}} &= \vec{E} + i \vec{B} = - \vec\nabla e^{-i \vec{a} \cdot \vec{\nabla}} \frac{Q}{R} = - \vec\nabla \frac{Q}{\sqrt{x^2 + y^2 + (z-ia)^2}} = - \vec\nabla Q \bar{f}(R)
\eld \label{eq:holoEMF}
\el
The funtions $f(R)$ and $\bar{f}(R)$ are harmonic. The components of the stress tensor eq.\eqc{eq:EMstress} are determined from electromagnetic fields eq.\eqc{eq:holoEMF} as follows.
\bl
\bld
T_{00} &= \frac{E^2 + B^2}{8 \pi} = \frac{H \cdot \bar{H} }{8 \pi} = Q^2 \frac{x^2 + y^2 + z^2 + a^2}{8 \pi \S^3}
\\ T_{0i} &= - \frac{1}{4 \pi} \left( \vec{E} \times \vec{B} \right)^i = - \frac{1}{8 \pi i} \left( \vec{H} \times \vec{\bar{H}} \right)^i = - Q^2 \frac{ \left( \vec{a} \times \vec{R} \right)^i }{4 \pi \S^3}
\\ T_{ij} &= \frac{- E^i E^j - B^i B^j + \delta_{ij} (E^2 + B^2) }{8 \pi} = \frac{- H^i \bar{H}^j - \bar{H}^i H^j + \delta_{ij} H \cdot \bar{H}}{8 \pi}
\\ &= Q^2 \left(\frac{- R^i R^j - a^i a^j}{4 \pi \S^3} + \delta_{ij} \frac{x^2 + y^2 + z^2 + a^2 }{8 \pi \S^3} \right)
\eld \label{eq:KNEMstress}
\el
The momentum space is defined as Fourier transform in 3-space; $\tilde{f}(\vec{q}) = \int f(\vec{r}) e^{- i \vec{q} \cdot \vec{r}} d^3r$. In momentum space, each component takes the following form\footnote{The Fourier transform was obtained by resumming the series expansion in $q^2 a^2$ and $\vec{q} \cdot \vec{a}$.}.
\bl
\bld
T_{00} &= - \frac{Q^2 \pi}{8} q J_0 (\vec{a} \times \vec{q})
\\ T_{0i} &= - \frac{i Q^2 \pi}{8} q \left[ J_1 (\vec{a} \times \vec{q}) \right]^i = - \frac{i Q^2 \pi}{8} q (\vec{a} \times \vec{q})^i \left[ \frac{J_1 (\vec{a} \times \vec{q}) }{\vec{a} \times \vec{q}} \right]
\\ T_{ij} &= \frac{Q^2 \pi}{8} q \left[ (\vec{a} \times \vec{q})^i (\vec{a} \times \vec{q})^j \right] \left[ \frac{J_2 (\vec{a} \times \vec{q})}{(\vec{a} \times \vec{q})^2} \right] + \frac{Q^2 \pi}{8} \frac{q^i q^j - q^2 \delta_{ij}}{q} \left[ \frac{J_1 (\vec{a} \times \vec{q}) }{\vec{a} \times \vec{q}} \right]
\eld
\el
The functions $\frac{J_n (\vec{a} \times \vec{q})}{(\vec{a} \times \vec{q})^n}$ are obtained by substituting powers of $x^2$ in $\frac{J_n (x)}{x^n}$ by powers of $(\vec{a} \times \vec{q})^2$. In the above expression $q^2$ should be understood as $(\vec{q})^2$. To covariantise the expression, let us introduce the following definitions.
\bg\label{EDef}
u^\m = \frac{P^\m}{m} = (1, \vec{0}) \,,\quad E^\m = - \frac{1}{m^2} \e^{\m\n\l\s}P_\n S_\l q_\s = (0, \vec{a} \times \vec{q}) \,.
\eg
This sets
\bl
\boxed{
\bld \label{eq:Stress_Tensor_Covariant_Form}
\frac{8 \, T_{\m\n}}{Q^2 \pi \sqrt{-q^2}} &= - u_\m u_\n J_0 (\vec{a} \times \vec{q}) + \left(- u_\m u_\n + \frac{q_\m q_\n - q^2 \eta_{\m \n}}{-q^2} + 2 i u_{(\m} E_{\n)} \right) \left[ \frac{J_1 (\vec{a} \times \vec{q}) }{\vec{a} \times \vec{q}} \right]
\\ &\phantom{=asdfasdf} + E_\m E_\n \left[ \frac{J_2 (\vec{a} \times \vec{q})}{(\vec{a} \times \vec{q})^2} \right]
\eld
}
\el
where $q^2 = \eta_{\m\n} q^\m q^\n = - ( \vec{q} )^2$. Note that the functions $\frac{J_n (\vec{a} \times \vec{q})}{(\vec{a} \times \vec{q})^n}$ can be expressed covariantly by modified Bessel's functions; $\frac{J_n (\vec{a} \times \vec{q})}{(\vec{a} \times \vec{q})^n} = \frac{I_n (E^\m) }{(E^\m)^n}$.
\bl
\frac{J_n(\vec{x})}{(\vec{x})^n} = \sum_{m=0}^\infty \frac{1}{2^n m!(m+n)!} \left(-\frac{\vec{x}^2}{4}\right)^{m} && \Rightarrow && \frac{I_n(x^\m)}{(x^\m)^n} = \sum_{m=0}^\infty \frac{1}{2^n m!(m+n)!} \left(\frac{x^\m x_\m}{4}\right)^{m} \,.
\el

\subsection{Matching with the metric}
The defining equation of general relativity, the Einstein's equations, relate the stress tensor to the Ricci tensor. Decomposing the metric $g_{\m\n} = \eta_{\m\n} + \k h_{\m\n}$ into the background metric $\eta_{\m\n}$ and metric perturbation $h_{\m\n}$ with $\k = \sqrt{32 \pi G}$, the Einstein's equation up to linear order in $h$ takes the form
\bl
\square h_{\m\n} + \p_\m \p_\n h - \p^\a \p_\m h_{\n\a} - \p^\a \p_\n h_{\m\a} &= \frac{\k}{2} P_{\m\n}^{\l\s} T_{\l\s} \label{eq:linEinsEq}
\el
where $h = \eta^{\m\n} h_{\m\n}$ is the trace of the metric perturbation and $P_{\m\n}^{\l\s} = \delta^{\l\s}_{\m\n} - \half \eta_{\m\n} \eta^{\l\s}$ is the trace-reverser. Adopting the de Donder gauge or the harmonic gauge $g^{\n\l} \G^\m_{\n\l} = 0$ simplifies this equation into
\bl
\square \bar{h}_{\m\n} &= \frac{\k}{2} T_{\m\n}
\el
where $\bar{h}_{\m\n} = P_{\m\n}^{\l\s} h_{\l\s}$ is the trace-reversed metric perturbation. If the stress tensor $T_{\m\n}$ is traceless as in eq.\eqc{eq:EMstress}, the trace-reverser in front of $h_{\m\n}$ can be dropped.
\bl
\eta^{\m\n} T_{\m\n} = 0 &\Rightarrow \square {h}_{\m\n} = \frac{\k}{2} T_{\m\n} \label{eq:TLEinstein}
\el
The metric up to this order can be obtained by solving the above equation. This has been demonstrated in~\cite{Donoghue:2001qc,Holstein:2006ud}.

The explicit form of harmonic coordinates for Kerr-Newman BHs has been obtained in~\cite{Lin:2014laa}, which has an all-order expansion in $G$ and $Q^2$. Since we are only interested in metric components of order $GQ^2$, we expand the explicit metric given in~\cite{Lin:2014laa} in powers of $G$ and $Q^2$ and then extract the wanted term. The result is;
\bl
\bld\label{eq:GQ2_Metric}
\frac{\left. ds^2 \right|_{GQ^2}}{GQ^2} &= \frac{ dt^2 }{\S} + \frac{ 2 a (y dx - x dy) dt }{\S (r^2 + a^2)} + \frac{ \left[ a ( y dx - x dy ) + r ( x dx + y dy + z (1 + \frac{a^2}{r^2}) dz ) \right]^2 }{\S (r^2 + a^2)^2} \,,
\eld 
\el
where $\left. ds^2 \right|_{GQ^2}$ denotes terms in the metric linear in $GQ^2$. Taking the d'Alembertian of metric perturbation components as in eq.\eqc{eq:TLEinstein} and comparing with the obtained stress tensor eq.\eqc{eq:KNEMstress} shows that the results are consistent.

\subsection{Matching $\mathcal{O}(GQ^2)$ impulse}
\begin{figure}[H]
\centering
\includegraphics[scale=0.4]{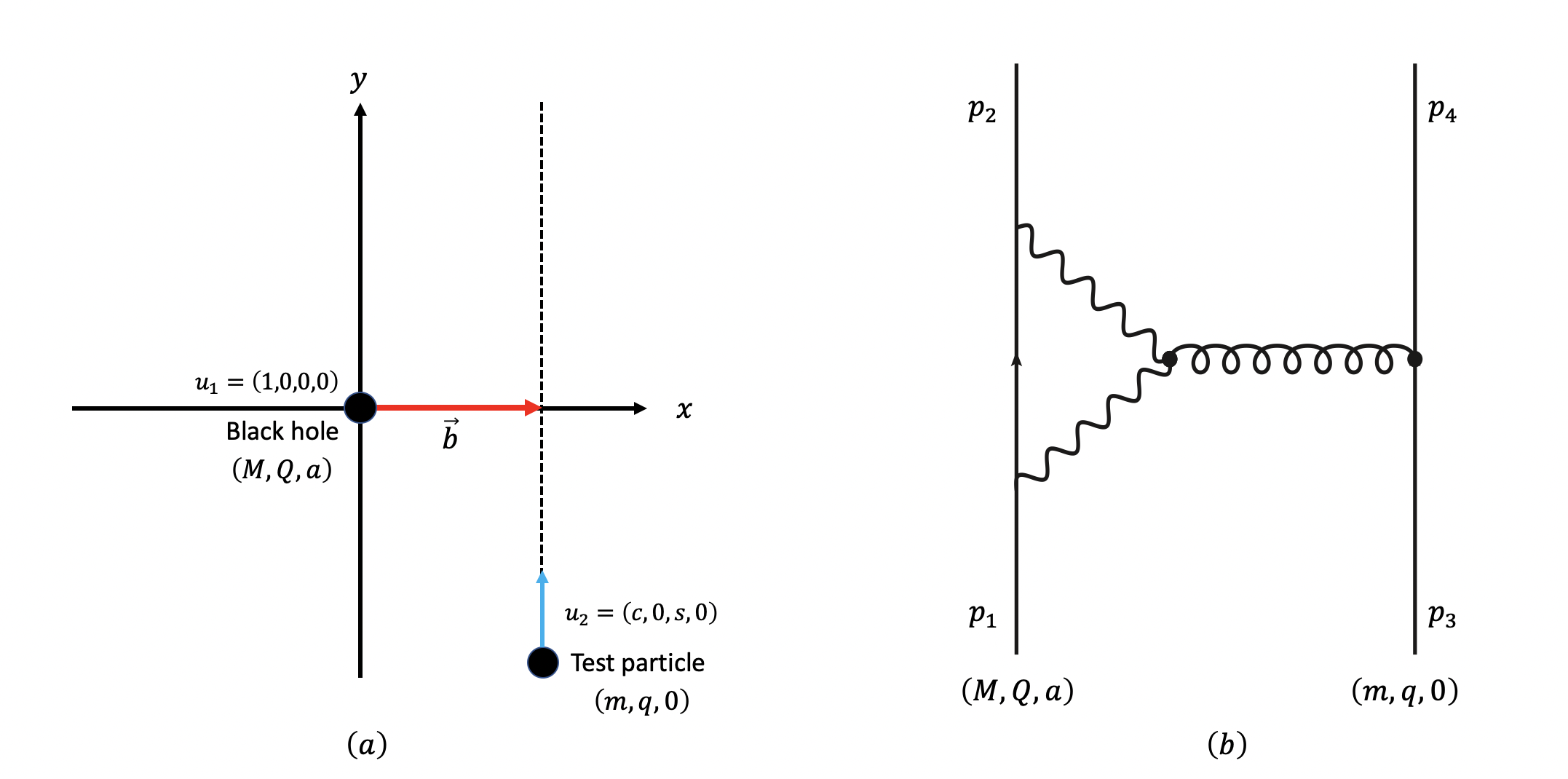}
\caption{(a) The scattering plane. The Kerr-Newman black hole of mass $M$, charge $Q$ and spin $a$ is centered at the origin. The scalar test particle of mass $m$ and charge $q$ is moving with initial proper velocity $u_1$ and the impact parameter is aliged with the $x$ axis. (b) The Feynman diagram needed to compute the impulse via amplitude method.}
\label{fig:Scattering_Plane2}
\end{figure}

In \citep{Chen:2021huj}, the impulse experienced by a charged scalar particle in Kerr-Newman background is computed to $\mathcal{O}(G)$ and $\mathcal{O}(qQ)$ by both classical equations of motion (Lorentz force equation + geodesic equation) and amplitude methods. It was shown there that the impusle computed from both methods are consistent with each other. In this subsection, we further extend the matching between the impulse from ampliutde and the geodesic equation to $\mathcal{O}(GQ^2)$ and confirm that the stress tensor eq.\eqref{eq:Stress_Tensor_Covariant_Form} we obtain is indeed correct.

\paragraph{Impulse from geodesic equation}

The impulse of $\mathcal{O}(GQ^2)$ can be computed from integrating the geodesic equation at constant velocity approximation:
\begin{equation}\label{eq:Geodesic_eq}
\Delta p_2^{\mu} \Big|_{GQ^2} = -m \int_{-\infty}^{\infty}d\tau \Gamma^{\mu}_{\nu\rho}\Big|_{GQ^2} u_2^{\nu}u_2^{\rho}.
\end{equation}
We set the scattering plane to be the $z=0$ plane and the Kerr-Newman black hole of mass $M$ sits at the origin of the scattering plane so its proper velocity is $u_1 = (1,0,0,0)$. For the test particle particle of mass $m$, the impact parameter is aligned with the $x$-axis and its proper velocity is: $u_2 = (\cosh w, 0, \sinh w, 0) \equiv (c_w, 0, s_w, 0)$, where $w$ is its rapidity. The Christoffel symbol in eq.\eqref{eq:Geodesic_eq} is computed from the Kerr Newman metric: 
\begin{equation}
\begin{split}
ds^2 &= -\rho^2 \left( \frac{dr^2}{\Delta} + d\theta^2 \right) + \left(dt -a\sin^2\theta \right)^2 - \frac{\sin^2\theta}{\rho^2}\left[(r^2 + a^2)d\phi - a dt\right]^2,\\
\Delta &= a^2 + \frac{GQ^2}{4\pi}  - 2 G M r + r^2,\\
\rho &= r^2 + a^2 \cos^2 \theta.
\end{split}
\end{equation}

Integrating eq.\eqref{eq:Geodesic_eq}, the impulse to all orders in the black hole's spin $a$ is computed to be:
\begin{equation}
\begin{split}
\Delta p^x &= \frac{ G Q^2 m \sinh w \left[2 a^3 \coth w- (b^2-a^2)^{3/2}-a^2 b \text{csch}^2 w-3 a^2 b+b^3\right]}{8 a^2 \left(b^2-a^2\right)^{3/2}}, \\
\Delta p^t &= \Delta p^y = \Delta p^z = 0.
\end{split}
\end{equation}
We will match with this equation order by order in spin $a$, so we expand it to $a^4$ for later convenience
\begin{equation}\label{eq:Impulse to a4}
\begin{split}
\frac{\Delta p^x}{GQ^2} &=
\frac{m \left(3 c_{2 w}+1\right)}{32 b^2 s_w}
-\frac{\left(m c_w\right)}{4 b^3}a
+\frac{3  m \left(7 c_{2 w}+1\right)}{128 b^4 s_w}a^2
\\
& \quad
-\frac{3  \left(m c_w\right)}{8b^5}a^3
+\frac{5  m \left(11 c_{2 w}+1\right)}{256 b^6 s_w} a^4+ \mathcal{O}(a^5).
\end{split}
\end{equation}

\paragraph{Impulse from amplitude}
It is known that the impulse can also be computed from amplitudes with the formula
\begin{equation}\label{eq:Impulse_from_Amp}
\Delta p^{\mu} = \frac{1}{4m Ms_w}\int \frac{d^2 \vec{
q}}{(2\pi)^2} e^{i\vec{q}\cdot\vec{b}} \times iq^{\mu}M_4.
\end{equation}
Since we are considering scalar particle in Kerr-Newman background, the relevant amplitude $M_4$ can be constructed by gluing the stress tensor we just obatined in eq.\eqref{eq:Stress_Tensor_Covariant_Form} with a graviton propagator and a scalar vertex together:
\begin{equation}\label{eq:4pt_Amplitude}
\begin{split}
M_4 &= T_{\mu\nu} \frac{iP^{\mu\nu\rho\sigma}}{q^2} V_{\rho\sigma}^{(\phi\phi h)}\\
&= \frac{i G Q^2 }{2} \frac{ M m^2}{\sqrt{-q^2}}\left[F_1 \cosh (2 w) + 4 i \frac{\mathcal{E}_a}{mM^2}\cosh (w) F_2+F_3  + \left(E_a^2 - \frac{2\mathcal{E}_a^2}{m^2M^4}\right) F_4 \right].
\end{split}
\end{equation}
The tensor $P^{\mu\nu\rho\sigma}$ in the graviton propagator and the scalar vertex are given by
\begin{equation}
\begin{split}
P^{\mu\nu\rho\sigma} &= \frac{1}{2}\left(\eta^{\mu\rho}\eta^{\nu\sigma} + \eta^{\nu\rho}\eta^{\mu\sigma} - \eta^{\mu\nu}\eta^{\rho\sigma} \right), \\
V_{\rho\sigma}^{(\phi\phi h)}\left(p_1, p_2\right) &= 
-\frac{i\kappa}{2}\left[ p_{1,\mu}p_{2,\nu} + p_{1,\nu}p_{2,\mu} - \eta_{\mu\nu}(p_1 \cdot p_2 - m^2) \right],
\\
\end{split}
\end{equation}
where we use the convention of incoming $p_1$ and out-going $p_2$, and we define the variables 
\begin{equation}
E_a^\mu = - \frac{1}{M^2} \e^{\m\n\l\s}P_\n S_\l q_\s = (0, 0, a q^x, 0), \;
\mathcal{E}_a = - M^2 p_3 \cdot E_a = -M^2 m s_w a q^x.
\end{equation}
The factors $F_1,\;F_2,\;F_3,\;F_4$ can be directly read off by comparining eq.\eqref{SpinBasis} and \eqref{eq:Stress_Tensor_Covariant_Form}. 

Now we are ready to compute the impulse by expanding eq.\eqref{eq:4pt_Amplitude} order by order in spin $a$ and Fourier transform it into the impact parameter space as in eq.\eqref{eq:Impulse_from_Amp}. We present the computation up to $a^4$ and we will see that the amplitude method indeed reproduces eq.\eqref{eq:Impulse to a4}:

\begin{itemize}[leftmargin=*]
\item $a^0$ \newline
We first start from the scalar piece. Only the leading term of $F_1$ and $F_3$ contributes here, giving
\begin{subequations}
\begin{equation}
M_4 \big|_{a^0} = -\frac{i\pi GQ^2 M m^2}{4\sqrt{-q^2}} \left(3c_{2w}+1\right).
\end{equation}
Using eq.\eqref{eq:Impulse_from_Amp}, 
\begin{equation}
\frac{\Delta p^x}{GQ^2} \Bigg|_{a^0} 
= \frac{1}{4Mms_w}\int\frac{d^2\vec{q}}{(2\pi)^2}e^{i\vec{q}\cdot \vec{b}} q^{x}  \frac{\pi M m^2}{4\sqrt{-q^2}} \left(3c_{2w}+1\right) 
= \frac{m \left(3 c_{2 w}+1\right)}{32 b^2 s_w},
\end{equation}
\end{subequations}
which is consistent with the first term in eq.\eqref{eq:Impulse to a4}.\footnote{
Here, we used the Fourier transform identity:
\begin{equation}
\int\frac{d^2\vec{q}}{(2\pi)^2}\frac{e^{i\vec{q}\cdot \vec{b}} }{\sqrt{-q^2}} \delta(u_1\cdot q)\delta(u_2\cdot q) =  \frac{1}{2\pi b s_w}. \nonumber
\end{equation}
}

\item $a^1$\newline
We move on to linear order in spin $a^1$. Here, only the $F_2$ term contributes:
\begin{subequations}
\begin{equation}
M_4 \big|_{a^1} = -\frac{\pi GQ^2 m}{M\sqrt{-q^2}} c_w\mathcal{E}_a = \frac{\pi GQ^2 M m^2}{\sqrt{-q^2}} s_w c_w a_a q^x.
\end{equation}
Using eq.\eqref{eq:Impulse_from_Amp}, 
\begin{equation}
\frac{\Delta p^x}{GQ^2} \Bigg|_{a^1} 
= \frac{\pi m}{4}c_w\int\frac{d^2\vec{q}}{(2\pi)^2}e^{i\vec{q}\cdot \vec{b}} \frac{(q^{x})^2}{\sqrt{-q^2}}
= -\frac{ \left(m_b c_w\right)}{4 b^3}a,
\end{equation}
\end{subequations}
which is also consistent with the second term in eq.\eqref{eq:Impulse to a4}.

\item $a^2$\newline
We move on to quadratic order in spin $a^2$. With the identities
\begin{equation}
\begin{split}
\mathcal{E}_a^2 &= -M^4 m^2  s_w^2 \left[ (q\cdot a)^2 - q^2 a^2 \right],\\
E_a^2 &= (q\cdot a)^2 - q^2 a^2,
\end{split}
\end{equation}
we can write the amplitude as
\begin{subequations}
\begin{equation}
\begin{split}
M_4\big|_{a^2} &= \frac{-i \pi GQ^2}{32} Mm^2 (5 c_{2w}+1)[q^2a^2 - (a\cdot q)^2] - \frac{i\pi GQ^2}{8M^3}\mathcal{E}_a^2 + \frac{i \pi GQ^2}{16} Mm^2 E_a^2 \\
&= \frac{-i \pi GQ^2}{32} Mm^2 (7 c_{2w}+1)[q^2a^2 - (a\cdot q)^2] .
\end{split}
\end{equation}
Using eq.\eqref{eq:Impulse_from_Amp}, the impulse at $a^2$ is computed to be
\begin{equation}
\Delta p^x \big|_{a^2} = \frac{3 m \left(7 c_{2 w}+1\right)}{128 b^4 s_w}a^2,
\end{equation}
\end{subequations}
and is again consistent with eq.\eqref{eq:Impulse to a4}.

\item $a^3$\newline
At cubic order in spin, the amplitude is
\begin{subequations}
\begin{equation}
M_4\big|_{a^3} = \frac{ GQ^2\pi}{8} mMc_w \mathcal{E}_a [q^2a^2 - (a\cdot q)^2] .
\end{equation}
The impulse at $a^3$ is computed to be
\begin{equation}
\Delta p^x \big|_{a^3} = -\frac{3  \left(m c_w\right)}{8b^5}a^3,
\end{equation}
\end{subequations}
and is again consistent with eq.\eqref{eq:Impulse to a4}.

\item $a^4$\newline
At quartic order in spin, the amplitude is
\begin{subequations}
\begin{equation}
M_4\big|_{a^4} = \frac{ i\pi GQ^2}{768} m^2 M(11c_{2w}+1) [q^2a^2 - (a\cdot q)^2]^2,
\end{equation}
and the impulse at $a^4$ is computed to be
\begin{equation}
\Delta p^x \big|_{a^4} = \frac{5  m \left(11 c_{2 w}+1\right)}{256 b^6 s_w}a^4.
\end{equation}
\end{subequations}
It is again consistent with eq.\eqref{eq:Impulse to a4}.
\end{itemize}
By matching to the impulse from classical EOM order by order in spin, we confirm our stress tensor eq.\eqref{eq:Stress_Tensor_Covariant_Form} to be correct.

\newpage
\section{Kerr-Newman stress tensor from form factors}
In this section, we extract the classical piece of the stress tensor at 1 PM via unitarity cuts, eq.\eqref{eq:On-shell_amplitude}.  We are considering form factors in momentum space with $q^2 \neq 0$, so that the stress tensor can be interpret as a three-point coupling involving a \emph{massive spin-2 particle}. Since we are interested in the spin-dependence for the stress-tensor, we will be computing the amplitude
\eq
M(1^s,2^s,q^2)=\epsilon_{\mu\nu}\langle p_1,s|T_{\mu\nu}(q)|p_2,s\rangle\,.
\eqe
Stripping off the massive polarization vector $\epsilon_{\mu\nu}$ then gives back the form factor. 

We are interested in charge dependent part of the form factor at 1 PM. This comes in as photon loops where the electromagentic minimal coupling exchanges photon with the Maxwell stress-tensor. As in \cite{Chung:2018kqs, Chung:2019duq, Guevara:2017csg}, the classical component only requires the calculation of the triangle coefficient, which can be captured by the triangle cut eq.\eqref{eq:On-shell_amplitude}. That is at $\mathcal{O}(Q^2)$, we have

\begin{equation}
\boxed{
\LAB{p_2, s}T^{Q^2}_{\mu\nu}(q)\RAB{p_1, s} = \frac{\Delta_{\mu\nu}(q, p_1, -p_2)}{2m}
}
\end{equation}
with 
\begin{equation}
\Delta_{\mu\nu}(q, p_1, -p_2) = \left. J^{-1} \times \text{LS} \times I_{\Delta} \right|_{q^2 \rightarrow 0}
\end{equation}
where $\text{LS}$ is defined in eq.\eqref{eq:LS}, $I_{\Delta}$ is the scalar triangle coefficient, $J^{-1}$ is the inverse Jacobian factor from solving the delta functions in eq.\eqref{eq:LS} and $q^2 \rightarrow 0$ indicates that we are taking the leading order in $q^2$ expansion. The overall factor of momentum conservation $\delta(p_2-p_1-q)$ has been dropped for simplicity. It is shown in \citep{Chung:2018kqs} that the $J^{-1}$ and $I_{\Delta}$ cancel with each other in leading order of $q^2$ expansion, so that 
\begin{equation}
\Delta_{\mu\nu}(q, p_1, -p_2) = \left. \text{LS} \right|_{q^2 \rightarrow 0}
\end{equation}
Here, we aim to capture the stress tensor in all orders in spins by evaluating the amplitude of finite spin particles in infinite spin limit. As argued in \cite{Chung:2019duq}, it is rather delicate to obtain a correct classical prediction from amplitudes of finite spins. Taking the infinite spin limit is an easy fix of the problem, which will ultimately lead to \eqref{eq:Stress_Tensor_Covariant_Form}.\footnote{Note that in principle one also needs to conduct Hilbert space matching, which can lead to modification to terms involving the spin-orbit coupling. However, for three-point kinematics in the infinite spin limit such terms are irrelevant. } 

\subsection{Kinematical set-up}
In this paper, we follow the parametrization of momenta introduced in \cite{Guevara:2017csg}:
\begin{equation}
\begin{split}
p_1 &= \left(E, -\vec{q}/2 \right) = \RAB{\lambda}\LSB{\eta} + \RAB{\eta}\LSB{\lambda}  \\
p_2 &= \left(E, +\vec{q}/2 \right) =\beta \RAB{\lambda}\LSB{\eta} + \frac{1}{\beta}\RAB{\eta}\LSB{\lambda} + \RAB{\lambda}\SB{\lambda}
\end{split}
\end{equation}
and the momentum transfer:
\begin{equation}
q = p_2 - p_1 = \left(0, \vec{q}\right) = \RAB{\lambda}\SB{\lambda} + \CO(\beta-1) = K + \CO(\beta-1)
\end{equation}
in the Holomorphic Classical Limit (HCL) introduced in \cite{Guevara:2017csg} where $q^2 = 0$ without $|\vec{q}| \rightarrow 0$. The triangle coefficient of eq.\eqref{eq:On-shell_amplitude} is captured by the following integral:
\begin{equation}\label{eq:LS}
\begin{split}
\text{LS}
&= \frac{1}{4} \sum_{h_1,h_2 = \pm 1}\int_{\Gamma_{LS}} d^4 L \delta(L^2-m^2) \delta(k_3^2) \delta(k_4^2) \\
&\qquad \qquad \qquad \times M_3(q, -k_3^{h_1}, -k_4^{h_2})\times M_3(1^s, k_3^{-h_1}, -L)\times M_3(-2^s, k_4^{-h_2}, L)\\
&=\sum_{h_1, h_2}\frac{\beta}{16(\beta^2-1)m^2} \int_{\Gamma_{LS}} \frac{dy}{y} M_3(q, -k_3^{h_1}, -k_4^{h_2})\times M_3(1^s, k_3^{-h_1}, -L^s) \times M_3(-2^s, k_4^{-h_2}, L^s)
\end{split}
\end{equation}

The two three-point amplitudes corresponding to Kerr-Newman BHs coupling to photons are minimally coupling three-point amplitudes in the infinite spin limit as pointed out in section \ref{subsec:Photon_Effective_Action}.
\begin{equation}
\begin{split}
M_3(-2^s, k_4^{\eta}, L^s) &=
x_2^{\eta} \oint \frac{dz}{2\pi i z} \left( \sum_{n=0}^\infty  z^n \right) \left( \frac{ [ \bf{2L} ] - \la \bf{2L} \ra }{2m} + \frac{\eta}{z} \frac{ \sbra{\bf{2}} k_4 \ket{\bf{L}} + \bra{\bf{2}} k_4 \sket{\bf{L}} }{4m^2} \right)^{2s} \\
M_3(1^s, k_3^{\eta}, -L^s) &=
x_1^{\eta} \oint \frac{dz}{2\pi i z} \left( \sum_{n=0}^\infty z^n \right) \left( \frac{ [ \bf{L1} ] - \la \bf{L1} \ra }{2m} + \frac{\eta}{z} \frac{ \sbra{\bf{L}} k_3 \ket{\bf{1}} + \bra{\bf{L}} k_3 \sket{\bf{1}} }{4m^2} \right)^{2s}
\end{split}
\end{equation}
where $\eta$ is the helicity of the massless particle and the $x$-factors are given by
\begin{equation}
x_4 = \frac{\sbra{k_4} L \ket{\z_1} }{\la k_4 \z_1 \ra m}, \quad 
x_3 = \frac{ \sbra{k_3} p_1 \ket{\z_2} }{\la k_3 \z_2 \ra m }
\end{equation}
The three-point amplitudes will take an exponential form in the infinite spin limit~\cite{Arkani-Hamed:2019ymq,Chung:2019duq}, and this property will also be shown to hold for the triangle cut.

The amplitude $M_3(q, -k_3^{h_1}, -k_4^{h_2})$ is uniquely fixed by three-point kinematics~\cite{Arkani-Hamed:2017jhn}:
\begin{equation}
M_3(q,-k_3^{+1},-k_4^{-1}) = - \frac{\lambda_{k_4}^4\SB{k_3 k_4}^{2}}{M_{pl} M^2}, \quad 
M_3(q,-k_3^{-1},-k_4^{+1}) = - \frac{\lambda_{k_3}^4\SB{k_3 k_4}^{2}}{M_{pl} M^2}
\end{equation}
where $M^2=q^2$. It is straight forward to demonstrate that the above is simply the on-shell form of the Maxwell stress tensor in the chiral basis. For our purposes, we want these amplitudes to be in the symmetric basis:
\begin{equation}
\begin{split}
M_3^{\text{sym}}(q,-k_3^{+1},-k_4^{-1}) 
&= - \frac{\lambda_{k_4, \a}\lambda_{k_4, \b} \lambda_{k_4}^{\g}\lambda_{k_4}^{\delta} \SB{k_3 k_4}^{2}}{M_{pl} M^2}\frac{q_{\gamma\dot{\a}}}{M}\frac{q_{\delta\dot{\b}}}{M}
= - \frac{\lambda_{k_4, \a} \tilde{\lambda}_{k_3,\dot{\a}} \lambda_{k_4, \b} \tilde{\lambda}_{k_3,\dot{\b}} }{M_{pl}}\\
&\rightarrow - \frac{\MixLeft{k_4}{\s_\m}{k_3}\MixLeft{k_4}{\s_\n}{k_3}}{M_{pl}}
\\
M_3^{\text{sym}}(q,-k_3^{-1},-k_4^{+1}) 
&= - \frac{\lambda_{k_3, \a}\lambda_{k_3, \b} \lambda_{k_3}^{\g}\lambda_{k_3}^{\delta}\SB{k_3 k_4}^{2}}{M_{pl} M^2} \frac{q_{\gamma\dot{\a}}}{M}\frac{q_{\delta\dot{\b}}}{M}
= -\frac{ \lambda_{k_3, \a}\tilde{\lambda}_{k_4,\dot{\a}} \lambda_{k_3, \b} \tilde{\lambda}_{k_4,\dot{\b}} }{M_{pl}}\\
&\rightarrow - \frac{\MixLeft{k_3}{\s_\m}{k_4}\MixLeft{k_3}{\s_\n}{k_4}}{M_{pl}}
\end{split}
\end{equation}

\subsection{Evaluating the integrand}
Combining all the intermediate results, the integrand can be separated into a spin-independent part and a spin-dependent part where wanted terms are picked out from the residue integral.
\begin{equation}
(M_3)^3 
= ( M_3^{s=0} )^3 \oint \frac{dz_1}{2 \pi i z_1} \frac{dz_2}{2 \pi i z_2} \left( \sum_{n=0}^\infty  z_1^n \right) \left( \sum_{n=0}^\infty  z_2^n \right) F (- \bf{2}, k_4, k_3, \bf{1})^{2s}\\ 
\end{equation}
The spin-independent part is given as
\begin{equation}
\begin{split}
( M_3^{s=0} )^3 &= m^2 \alpha_q^2 \left( \frac{\sbra{k_4} L \ket{\z_1} }{\la k_4 \z_1 \ra m} \right)^{\eta_1} \left( \frac{ \sbra{k_3} p_1 \ket{\z_2} }{\la k_3 \z_2 \ra m } \right)^{\eta_2} M_3 (- k_4^{- \eta_1}, - k_3^{- \eta_2}, \bf{q})\\
\end{split}
\end{equation}
with 
\begin{equation}
\begin{split}
M_3 (- k_4^{+}, - k_3^{-}, \bf{q}) &= - \frac{\bra{k_3} \s_\m \sket{k_4} \bra{k_3} \s_\n \sket{k_4}}{M_{pl}}\\
M_3 (- k_4^{-}, - k_3^{+}, \bf{q}) &= - \frac{\bra{k_4} \s_\m \sket{k_3} \bra{k_4} \s_\n \sket{k_3}}{M_{pl}}
\end{split}
\end{equation}
The residue integrand for the spin-dependent part is
\begin{equation}
\begin{split}
F (- \bf{2}, k_4, k_3, \bf{1}) &= \left( \frac{ [ \bf{2L} ] - \la \bf{2L} \ra }{2m} + \frac{\eta_1}{z_1} \frac{ \sbra{\bf{2}} k_4 \ket{\bf{L}} + \bra{\bf{2}} k_4 \sket{\bf{L}} }{4m^2} \right) \times \left( \frac{ [ \bf{L1} ] - \la \bf{L1} \ra }{2m} + \frac{\eta_2}{z_2} \frac{ \sbra{\bf{L}} k_3 \ket{\bf{1}} + \bra{\bf{L}} k_3 \sket{\bf{1}} }{4m^2} \right)\\
&= \bar{u}(2) u(1) - \left( \frac{\eta_1}{z_1} k_4^\m + \frac{\eta_2}{z_2} k_3^\m \right) S^\m_{1/2}
 + \frac{k_3^\m}{2m} \left( \frac{p_1^\m + p_2^\m}{2m} \bar{u}(2) u(1) - \frac{q^\n}{2m} \bar{u}(2) \g_{\m\n} u(1) \right)
\\ &\phantom{=asd} - \left( \frac{\eta_1}{z_1} + \frac{\eta_2}{z_2} \right) \left( \frac{q^2}{8m^2} \bar{u}(2) \g^5 u(1) + \frac{k_4^\m k_3^\n}{4m^2} \bar{u}(2) \g_{\m\n} \g^5 u(1) \right)
\\ &\phantom{=asd} - \frac{\eta_1 \eta_2}{z_1 z_2} \left( \frac{q^2}{8m^2} \bar{u}(2) u(1) + \frac{k_4^\m k_3^\n}{4m^2} \bar{u}(2) \g_{\m\n} u(1) \right)
\end{split}
\end{equation}
\noindent
where $\g^5 = i \g^0 \g^1 \g^2 \g^3$ and $\g^{\m\n} = \half [\g^\m, \g^\n]$. The definition for spin-$\half$ spin vector $S^\m_{1/2}$ is given as
\begin{equation}\label{eq:Spin vec def}
\begin{split}
\bar{u}(p_2) u(p_1) &= \frac{ [ \bf{21} ] - \la \bf{21} \ra }{2m}\\
S_{1/2}^\m &= \half \bar{u}(p_2) \g^\m \g_5 u(p_1) = - \frac{1}{4m} \left( \sbra{\bf{2}} \bar{\s}^\m \ket{\bf{1}} + \bra{\bf{2}} \s^\m \sket{\bf{1}} \right)
\end{split}
\end{equation}
In the HCL only the first two terms of $F$ survive:
\begin{equation}
\begin{split}
F 
&\stackrel{HCL}{\to} 1 - \left( \frac{\eta_1}{z_1} k_4^\m + \frac{\eta_2}{z_2} k_3^\m \right) \frac{S_{1/2}^\m }{m} \\
&= 1 - \eta \left( \frac{k_4^\m}{z_1}  - \frac{k_3^\m}{z_2}  \right) \frac{S_{1/2}^\m }{m} 
= 1 + \eta \left[ \frac{(y-1)^2}{4 z_1 y} + \frac{(y+1)^2}{4 z_2 y} \right] \frac{K \cdot S_{1/2}}{m} + \CO(\b-1)
\end{split}
\end{equation}
where $\eta = \eta_1$ is defined as the helicity of the $k_4$ photon on the 2 massive amplitude leg. Going to $s \to \infty$ limit and adopting the definition $\frac{1}{2s} S^\m = S^\m_{1/2}$, the $F$ term can be written as
\bl
\lim_{s\rightarrow \infty}F^{2s} = \exp{\left\lbrace \eta \left[ \frac{(y-1)^2}{4 z_1 y} + \frac{(y+1)^2}{4 z_2 y} \right] \frac{K \cdot S}{m} \right\rbrace} \equiv \exp\left[\eta f\times (a\cdot q)\right]
\el
where 
\begin{equation}
f \equiv \frac{(y-1)^2}{4 z_1 y} + \frac{(y+1)^2}{4 z_2 y}
\end{equation}
Summarising, the integrand becomes
\bl\label{eq:Partial Integrand}
(M_3)^3 &\stackrel{HCL}{\to} ( M_3^{s=0} )^3 \oint dZ \exp\left[\eta f\times (a\cdot q)\right]
\el
where we adopted the definition
\bl
\oint dZ \equiv \oint \frac{dz_1}{2 \pi i z_1} \frac{dz_2}{2 \pi i z_2} \left( \sum_{n=0}^\infty  z_1^n \right) \left( \sum_{n=0}^\infty z_2^n \right)
\el
for simplicity.

\subsection{Comparison with classical computations}
In section \ref{sec:Intro}, we provided a basis for expressing the stress tensor form factor, so
\begin{equation}
\Delta_{\mu\nu} = 2m\LAB{p_2}T_{\mu \nu} \RAB{p_1} = F_1 P_{\mu}P_{\nu} + 2 F_2 P_{(\mu}E_{\nu)} + F_3(q_{\mu}q_{\nu} - \eta_{\mu\nu} q^2) + F_4 E_{\mu}E_{\nu}
\end{equation}
We can extract the $F_1$, $F_2$, $F_3$, $F_4$ coefficients by contracting the integrand with different vectors or tensors to get a set of linear combinations of these four coefficients. By solving the equations, we show that the solution is consistent with the HCL version of eq.\eqref{eq:Stress_Tensor_Covariant_Form} where $(\vec{a} \times \vec{q})^2 = |\vec{a}|^2|\vec{q}|^2 - (\vec{a} \cdot \vec{q})^2 \stackrel{\text{HCL}}{\rightarrow} - (\vec{a} \cdot \vec{q})^2$. In this limit, the Bessel's function expressions of \eqref{eq:Stress_Tensor_Covariant_Form} are converted to modified Bessel's function expressions.
\bl
\frac{J_n (\vec{a} \times \vec{q})}{(\vec{a} \times \vec{q})^n} \stackrel{\text{HCL}}{\to} \frac{I_n (a \cdot q)}{(a \cdot q)^n}
\el

The kinematical set-up we are using implies the conditions
\begin{equation}
P\cdot q = P \cdot E = 0
\end{equation}
so that we will be solving the following system of linear equations:
\begin{equation}\label{eq:Linear equations}
\begin{split}
\Delta_{\mu\nu}P^\m P^\n &= m^4 F_1 - m^2 q^2 F_3 \\
\Delta_{\mu\nu}P^\m E^\n &= m^2 (\vec{a}\cdot \vec{q})^2 F_2 \\
\Delta_{\mu\nu}E^\m E^\n &= (\vec{a} \cdot \vec{q})^2 \left(F_4 (\vec{a}\cdot\vec{q})^2 - q^2 F_3\right) \\
\Delta_{\mu\nu} \eta^{\mu\nu} &= m^2 F_1 - 3q^2 F_3 + (\vec{a} \cdot \vec{q})^2 F_4 = 0
\end{split}
\end{equation}
where the last line is the traceless condition of the stress tensor. In the following sections, we explicitly show the computations of $\Delta_{\mu\nu}P^\m P^\n$, $\Delta_{\mu\nu}E^\m P^\n$ and $\Delta_{\mu\nu}E^\m E^\n$.

\paragraph{Computing $\Delta_{\m\n} P^\m P^\n$}$\phantom{123}$\newline
Contracting the stress tensor with $P_{\mu}P_{\nu}$  can help us read out  out the terms proportional to $P_\m P_\n$ and the $\eta_{\mu\nu}q^2$ part
\begin{equation}\label{eq:AngPSqr}
\begin{split}
\bra{k_3} P \sket{k_4} &= -( \b - 1 ) m^2\\
\bra{k_4} P \sket{k_3} &= -\frac{\b - 1}{\b} m^2
\end{split}
\end{equation}
where $P = (1-\a) P_3 + \a P_4$. Note that the above results are independent of $\a$\footnote{Overall factor of $\frac{\alpha_q^2}{M_{pl}}$ temporarily suppressed}. Therefore in the HCL the integrand will take the form
\begin{equation}
\begin{split}
(M_3)^3 
= 
2 m^6 (\b - 1)^2 \oint dZ \cosh \left[f\times (a\cdot q)\right] + \CO(\beta-1)^3
\end{split}
\end{equation}
To compare with the classical computation eq.\eqref{eq:Stress_Tensor_Covariant_Form}, we need to expand the integrand as a series in $\left(a \cdot q \right)^{2}$:
\begin{equation}
\begin{split}
\Delta_{\mu\nu}P^{\mu}P^{\nu} 
&= \frac{\b}{16m^2 (\beta^2 - 1)} \int\frac{dy}{y} (M_3)^{3} 
= \frac{(\b-1)  \b m^4}{8 (\b+1)}  \sum_{n=0}^{\infty} \int \frac{dy}{y}  	\oint dZ \frac{f^{2n}}{(2n)!}\left(a \cdot q \right)^{2n}\\
&= - \frac{(\b-1) \b m^4}{8 (\b+1)} \sum_{n=0} \frac{1}{2^{2n}(n!)^2} \left(a \cdot q \right)^{2n}
= - \frac{(\b-1) \b m^4}{8 (\b+1)} I_0 (a \cdot q)
\end{split}
\end{equation}
where $I_0$ is the modified Bessel's functions.

\paragraph{Computing $\Delta_{\m\n} P^\m E^\n$}$\phantom{123}$\newline
The vector $E_{\mu}$ can be spanned in the $\mathcal{U}_{\mu}$, $v_{\mu}$, $K_{\mu}$, $R_{\mu}$ basis defined in Appendix \ref{sec:Param}
\begin{equation}
E^{\mu} = \frac{i}{m^2} \left[\left(\mathcal{U}^{\mu} - v^{\mu} + \frac{\b^2-1}{\b}R^{\mu} \right)(K\cdot S) - K^{\mu} S \cdot \left( u-v + \frac{\b^2-1}{\b}R \right)\right]
\end{equation}
Contracting $E_{\mu}$ with $\MixLeft{k_4}{\sigma^{\mu}}{k_3}$ and $\MixLeft{k_3}{\sigma^{\mu}}{k_4}$ yields
\begin{equation}\label{eq:AngESqr}
\MixLeft{k_4}{E}{k_3} =  \frac{i  \left(y^2+1\right)}{2 y} (q\cdot S) (\beta -1) + \CO(\beta-1)^2 = -\MixLeft{k_3}{E}{k_4} 
\end{equation}
Combining with eq.\eqref{eq:AngPSqr}, the integrand  in the HCL will take the form
\begin{equation}
\begin{split}
(M_3)^3 
&=
-\frac{i  m^5 \left(y^2+1\right)(q\cdot a)}{y}  (\beta -1)^2  \oint dZ\sinh \left[ f \times (q\cdot a) \right]
\end{split}
\end{equation}
Evaluating the $y$, $z_1$ and $z_2$ integrals then:
\begin{equation}
\begin{split}
\Delta_{\mu\nu}P^{\mu}E^{\nu} 
& = \frac{\b}{16m^2(\beta^2-1)}\int \frac{dy}{y}(M_3)^3  \\
&= \frac{-i  m^3 \beta (\beta -1)}{16(\beta + 1)} \int \frac{dy}{y}\frac{ \left(y^2+1\right)}{y}   \sum_{n=0}^{\infty} \oint dZ  \frac{f^{2n+1}}{(2n+1)!}(q\cdot a)^{2n+2}\\
&= \frac{im^3 \b (\b - 1)}{8(\b + 1)} \sum_{n=0}^{\infty} \frac{ (a\cdot q)^{2n+2} }{2^{2n+1}n!(n+1)!}
= \frac{im^3 \b (\b - 1)}{8(\b + 1)} (a\cdot q)I_1 (a \cdot q)
\end{split}
\end{equation}
Solving for $F_2$ yields
\begin{equation}
F_2 
= \frac{i m \b (\b - 1)}{8(\b + 1)} \sum_{n=0}^{\infty} \frac{(a\cdot q)^{2n}}{2^{2n+1}n!(n+1)!}  
= \frac{i m \b (\b - 1)}{8(\b + 1)} \frac{I_1 (a \cdot q)}{a \cdot q}
\end{equation}

\paragraph{Computing $\Delta_{\m\n} E^\m E^\n$}$\phantom{123}$\newline
From eq.\eqref{eq:AngESqr}
\begin{equation}
\MixLeft{k_4}{E}{k_3}^2 = \MixLeft{k_3}{E}{k_4}^2  = -\frac{ \left(y^2+1\right)^2}{4 y^2} (q\cdot S)^2 (\beta -1)^2 + \CO(\beta-1)^3
\end{equation}
The integral becomes
\begin{equation}
\begin{split}
\Delta_{\m\n} E^\m E^\n 
& = \frac{\b}{16m^2(\beta^2-1)}\int \frac{dy}{y}(M_3)^3  \\
&= - \frac{\beta(\beta-1)m^2}{32(\beta+1)}(q \cdot a)^2 \int \frac{dy}{y} \frac{\left(y^2+1\right)^2}{ y^2}  \oint dZ\cosh\left[f\times (a\cdot q)\right]\\
&=  \frac{\beta(\beta-1)m^2}{16(\beta+1)} \sum_{n=0}^{\infty} \frac{2n+1}{2^{2n}n!(n+1)!}(a \cdot q)^{2n+2}\\
&= \frac{\beta(\beta-1)m^2}{8(\beta+1)} \left[ (a\cdot q)^2 I_2(a\cdot q) + (a\cdot q) I_1(a\cdot q)\right]
\end{split}
\end{equation}

\paragraph{Solving for the coefficients $F_1$, $F_2$, $F_3$, $F_4$}$\phantom{123}$\newline
The solution for eq.\eqref{eq:Linear equations} is 
\begin{itemize}[leftmargin=*]
\item $F_1$
\begin{equation}
\begin{split}
F_1 &=  \frac{1}{m^2}\frac{\Delta_{\mu\nu}E^{\mu}E^{\nu} }{(a\cdot q)^2}+ \frac{2}{m^4} \Delta_{\mu\nu}P^{\mu}P^{\nu}
= -\frac{1}{m^2} \frac{(\beta - 1)\beta m^2 }{8(\beta+1)}  \left[ \frac{I_1(a\cdot q)}{(a\cdot q)} + I_0(a\cdot q) \right]
\end{split}
\end{equation}
\item $F_2$
\begin{equation}
\begin{split}
F_2 
= \frac{i m^2 \b (\b - 1)}{8(\b + 1)} \frac{1}{m} \frac{I_1 (a \cdot q)}{a \cdot q}
\end{split}
\end{equation}
\item $F_3$
\begin{equation}
\begin{split}
F_3 &= \frac{1}{q^2}  \left( \frac{\Delta_{\mu\nu}E^{\mu}E^{\nu}}{ (a\cdot q)^2} + \frac{\Delta_{\mu\nu}P^{\mu}P^{\nu}}{m^2}\right) 
= - \frac{1}{q^2} \frac{(\beta - 1)\beta m^2 }{8(\beta+1)} \frac{I_1(a\cdot q)}{(a\cdot q)}
\end{split}
\end{equation}
\item $F_4$
\begin{equation}
\begin{split}
F_4 &=  \frac{2 \Delta_{\mu\nu}E^{\mu}E^{\nu} }{(a\cdot q)^4}+ \frac{1}{m^2} \frac{\Delta_{\mu\nu}P^{\mu}P^{\nu}}{(a\cdot q)^2} = \frac{(\beta - 1)\beta m^2 }{8(\beta+1)} \frac{I_2(a\cdot q)}{(a \cdot q)^2}
\end{split}
\end{equation}

\end{itemize}
In summary, the form factor from calculating $\Delta_{\mu\nu}$ is 
\begin{equation}
\boxed{
\begin{split}
\LAB{p_2}T_{\mu\nu} \RAB{p_1}
=
\frac{|\vec{q}| }{32} \frac{\alpha_q^2}{M_{pl}}
&\left\lbrace 
- \frac{P_{\mu}P_{\nu}}{m^2}\left[\frac{I_1(a \cdot q)}{(a \cdot q)} + I_0(a \cdot q)\right] 
+ 2i\frac{P_{(\mu}E_{\nu)}}{m}\frac{I_1(a \cdot q)}{(a \cdot q)} \right.\\
& \left.
\quad - \frac{q_{\mu}q_{\nu} - \eta_{\mu\nu}q^2}{q^2}\frac{I_1(a \cdot q)}{(a \cdot q)}
+ E_{\mu}E_{\nu}\frac{I_2(a \cdot q)}{(a \cdot q)^2} 
\right\rbrace
\end{split}
}
\end{equation}
which indeed reproduces the HCL version of eq.\eqref{eq:Stress_Tensor_Covariant_Form}.

\section{Conclusions}
In this paper, via computing the 1 PM $\mathcal{O}(Q^2)$ term of the stress-tensor, we demonstrate that the electro-magnetic minimal coupling for spinning particles can be mapped to the electro-magnetic dynamics of Kerr-Newman black hole to all order in spin. In the process, we've introduced a convenient basis to parameterize the stress-tensor form factor to all orders in spin variables, i.e. eq.(\ref{SpinBasis}). The $\mathcal{O}(Q^2)$ contribution arrises from one-loop photon exchanges, leading to triangle integrals that yield classical contributions. We take the classical-spin limit directly in the cut, to greatly simplify the computation.

For non-black hole charged sources, our methods apply, where one simply allows for deviation of the minimal coupling. Then the corresponding $\mathcal{O}(Q^2)$ modifications of the stress-tensor at 1 PM can be straight forwardly computed along the same lines. 

\section{Acknowledgements}
The authors would like to thank Sangmin Lee for helpful discussions.
 MZC and YTH is supported by MoST Grant No. 106-2628-M-002-012-MY3. YTH is also supported by Golden Jade fellowship. The work of JWK  was supported in part by the National Research Foundation of Korea (NRF) Grant 2016R1D1A1B03935179. JWK was also supported in part by Kwanjeong Educational Foundation.

\appendix
\section{Parametrization}\label{sec:Param}
The parametrization of the momenta is
\begin{equation}
\begin{split}
p_1 &= \RAB{\lambda} \LSB{\eta} + \RAB{\eta} \LSB{\lambda}\\
p_2 &= \beta \RAB{\lambda} \LSB{\eta} + \frac{1}{\beta} \RAB{\eta} \LSB{\lambda} + \RAB{\lambda} \LSB{\lambda}
\end{split}
\end{equation}
and the loop momenta $k_3$ and $k_4$ are 
\begin{equation}
\begin{split}
k_3 &=
\frac{1}{\beta + 1} \left( \RAB{\eta}(\beta^2-1)y - \frac{1}{y} \RAB{\lambda}(1+\beta y) \right)
\times
\frac{-1}{\beta + 1} \left( \LSB{\eta}(\beta^2-1)y + \LSB{\lambda}(1+\beta y) \vphantom{\frac{1}{y}}\right)\\
k_4 &=
\frac{1}{\beta + 1} \left( \frac{1}{\beta} \RAB{\eta}(\beta^2-1) + \frac{1}{y} \RAB{\lambda}(1 - y) \right)
\times
\frac{-1}{\beta + 1} \left( - \beta \LSB{\eta}(\beta^2-1)y + \LSB{\lambda}(1 - \beta^2 y) \vphantom{\frac{1}{y}}\right)
\end{split}
\end{equation}
For convenience, we introduce the basis vectors
\begin{equation}
\begin{split}
\mathcal{U}^{\mu} &= \frac{1}{2}\MixLeft{\eta}{\sigma^{\mu}}{\lambda}, v^{\mu} = \frac{1}{2}\MixLeft{\lambda}{\sigma^{\mu}}{\eta} \\
K^{\mu} &= \frac{1}{2}\MixLeft{\lambda}{\sigma^{\mu}}{\lambda}, R^{\mu} = \frac{1}{2}\MixLeft{\eta}{\sigma^{\mu}}{\eta} 
\end{split}
\end{equation}
So that the $x$ factor are paramaterized by
\begin{equation}
\begin{split}
x_4 &= \frac{\MixLeft{\z_1}{ L}{k_4} }{\AB{ k_4 \z_1 }m} = - y \b \\
x_3 &= \frac{\MixLeft{\z_2}{ p_1}{k_3}  }{\AB{ k_3 \z_2 }m} = + y \\
\end{split}
\end{equation}
and the vectors in the spin$2$-$\g$-$\g$ leg are paramaterized by
\begin{equation}
\begin{split}
-\frac{1}{2} \MixLeft{k_3}{\s_\m}{k_4} &= -\frac{(\beta -1) \left(\beta ^2 y-1\right)}{\beta +1} u_\m + \frac{(\beta -1) \beta (\beta  y+1)}{\beta +1} v_\m\\ &\phantom{=asdf} + \frac{(\beta  y+1) \left(\beta ^2 y-1\right)}{(\beta +1)^2 y} K_\m - (\beta -1)^2 \beta  y R_\m\\
-\frac{1}{2} \MixLeft{k_4}{\s_\m}{k_3} &= \frac{(\beta -1) (\beta  y+1)}{\beta  (\beta +1)} u_\m - \frac{(\beta -1) (y-1)}{\beta +1} v_\m
\\ &\phantom{=asdf} - \frac{(y-1) (\beta  y+1)}{(\beta +1)^2 y} K_\m + \frac{(\beta -1)^2 y}{\beta } R_\m
\end{split}
\end{equation}

\section{An alternative way to evaluate the integrand}
In \citep{Arkani-Hamed:2019ymq}, it is shown that the minimal coupling amplitude with two massive legs takes an exponential form in the infinite spin limit:
\begin{equation}
\begin{split}
M_3(-2^s, k_4^{\eta}, L^s) &=
\alpha_q m x_4^{\eta}\exp\left(-\eta k_4 \cdot a_2\right) \\
M_3(1^s, k_3^{\eta}, -L^s) &=
\alpha_q mx_3^{\eta} \exp\left(-\eta k_3 \cdot a_1\right) 
\end{split}
\end{equation}
where $a_1$ and $a_2$ are defined as
\begin{equation}\label{eq:Spin vec def 2}
\begin{split}
a_1 &= \frac{S}{m^2}\MixLeft{\BS{1}}{\sigma^{\mu}}{\BS{1}} \\
a_2 &= \frac{S}{m^2}\MixLeft{\BS{L}}{\sigma^{\mu}}{\BS{L}}
\end{split}
\end{equation}
Note that $a_1$ and $a_2$ are operators acting on different little group spaces. We will show that the integrand becomes
\begin{equation}
\begin{split}
\left(M_3\right)^3 =
- \frac{\alpha_q^2 m^2}{M_{pl}} 
&\times
\left\lbrace
\bra{k_4} \s_\m \sket{k_3} \bra{k_4} \s_\n \sket{k_3} \left(\frac{x_4}{x_3}\right) e^{-(k_4 - k_3)\cdot a}
\right. \\
&\left.
 \qquad\qquad
+\bra{k_3} \s_\m \sket{k_4} \bra{k_3} \s_\n \sket{k_4} \left(\frac{x_3}{x_4}\right) e^{+(k_4 - k_3)\cdot a}
\right\rbrace
\end{split}
\end{equation}
in the HCL which corresponds to eq.(\ref{eq:Partial Integrand}) after integrating out the $Z$ integral.

When gluing the massive lines together, the exponent in the exponential function takes the form:
\begin{equation}
\left(k_3 \cdot s_1 - k_4 \cdot s_2\right) + \text{commutator}= \frac{S}{m}\left(\MixLeft{\BS{1}}{k_3}{\BS{1}} - \MixLeft{\BS{L}}{k_4}{\BS{L}}\right)+ \text{commutator}
\end{equation} 
where the commutator term comes from the BCH formula. We first show that $k_4 \cdot a_2 = k_4 \cdot a_1 + \mathcal{O}(q^2)$ when boosting the spinors $\LAB{\BS{L}}$ and $\RSB{\BS{L}}$ to the $p_1$ frame and then that the commutators are at least of order $q^2$.

The boost operator on the massive spinors are worked out in \citep{Arkani-Hamed:2019ymq}
\begin{equation}\label{eq:Boost}
\begin{split}
\RAB{\BS{L}} &= \RAB{\BS{1}} + \frac{1}{m}k_3\RSB{\BS{1}} \Rightarrow
\LAB{\BS{L}} = \LAB{\BS{1}} - \frac{1}{m^2}\LSB{\BS{1}} \bar{k}_3 \\
\RSB{\BS{L}} &= \RSB{\BS{1}} + \frac{1}{m}k_3\RAB{\BS{1}}
\end{split}
\end{equation}
So that $\MixLeft{\BS{L}}{k_4}{\BS{L}}$ becomes
\begin{equation}
\MixLeft{\BS{L}}{k_4}{\BS{L}} = 
\MixLeft{\BS{1}}{k_4}{\BS{1}} 
- \frac{1}{m}\SB{\BS{1}|\bar{k}_3k_4|\BS{1}} 
+ \frac{1}{m}\AB{\BS{1}|k_4\bar{k}_3|\BS{1}}
- \frac{1}{m^2} \LSB{\BS{1}}\bar{k}_3k_4\bar{k}_3\RAB{\BS{1}}
\end{equation}
With the identities
\begin{equation}
\begin{split}
\SB{\BS{1}|\bar{k}_3k_4|\BS{1}} 
&= -\frac{2p_1 \cdot k_3}{m} \MixLeft{\BS{1}}{k_4}{\BS{1}} + \frac{2 p_1 \cdot k_4}{m} \MixLeft{\BS{1}}{k_3}{\BS{1}} - \AB{\BS{1}|k_3 \bar{k}_4|\BS{1}}\\
&=  -\frac{q^2}{m} \MixLeft{\BS{1}}{k_3}{\BS{1}} - \AB{\BS{1}|k_3 \bar{k}_4|\BS{1}}\\
\LSB{\BS{1}}\bar{k}_3k_4\bar{k}_3\RAB{\BS{1}} & = q^2 \MixLeft{\BS{1}}{k_3}{\BS{1}}
\end{split}
\end{equation}
everything combines into
\begin{equation}
\begin{split}
\MixLeft{\BS{L}}{k_4}{\BS{L}} = \MixLeft{\BS{1}}{k_4}{\BS{1}} + \frac{1}{m} \left(\AB{\BS{1}|k_4\bar{k}_3|\BS{1}} + \AB{\BS{1}|k_3\bar{k}_4|\BS{1}}\right) = \MixLeft{\BS{1}}{k_4}{\BS{1}} + \frac{q^2}{m^2} \AB{\BS{11}}
\end{split}
\end{equation}

Now, we turn to the commutator term where
\begin{equation}
\text{commutator} = \MixLeft{\BS{1}}{k_3}{\BS{1}^I}\MixLeft{\BS{L}_I}{k_4}{\BS{L}} - \MixLeft{\BS{L}}{k_4}{\BS{L}^I}\MixLeft{\BS{1}_I}{k_3}{\BS{1}}  + \cdots
\end{equation}
where the ellipsis corresponds to multiple commutator terms. Inserting the boost relation eq.\eqref{eq:Boost}, the commutator term becomes
\begin{equation}
\begin{split}
&
\left(
-\MixLeft{\BS{1}}{k_3 \bar{p}_1k_4}{\BS{L}} 
- \frac{1}{m} \MixLeft{\BS{1}}{k_3}{\BS{1}^I}\SB{\BS{1}_I|\bar{k}_3 k_4|\BS{1}}
\right)
-\left(
-\MixLeft{\BS{L}}{k_4 \bar{p}_1k_3}{\BS{1}}
+ \frac{1}{m} \AB{\BS{L}|k_4 k_3|\BS{1}^I} \MixLeft{\BS{1}_I}{k_3}{\BS{1}}
\right)
\\
=& m q^2 \left( \AB{\BS{11}} - \frac{1}{m} \MixLeft{\BS{1}}{k_3}{\BS{1}} \right)
\end{split}
\end{equation}
So that the commutators are at least of $q^2$ order. 
\begin{equation}
k_4 - k_3 = - \left( \frac{1+y^2}{2y} \right) K + \mathcal{O}(\beta - 1) \equiv \tilde{f} \times K+ \mathcal{O}(\beta-1)
\end{equation}
Summarizing:
\begin{equation}
e^{-\eta k_4\cdot a_2}e^{\eta k_3\cdot a_1} 
= e^{ - \eta \left(k_4 - k_3\right)\cdot a_1 + \mathcal{O}(q^2) } 
= e^{ \frac{1+y^2}{2y} K \cdot a_1 + \mathcal{O}(q^2) }
\end{equation}
where the definition for the spin vector $s_1$ in eq.\eqref{eq:Spin vec def 2} is off by the one in eq.\eqref{eq:Spin vec def} by $O(q^2)$ order. Carrying out the $dZ$ integral in equation eq.\eqref{eq:Partial Integrand}, one finds 
\begin{equation}
\oint dZ \exp\left[\eta f\times (a\cdot q)\right] = e^{ \frac{1+y^2}{2y} q \cdot a}
\end{equation}
Which is an exact match!!

%


\bibliography{mybib}{}
\bibliographystyle{JHEP}

\end{document}